\documentstyle[aps,prd,eqsecnum]{revtex}
\draft
\begin{document}

\title{Zero-point energies and the multiplicative anomaly.}

\author{J.~J.~McKenzie-Smith\thanks{email: 
{\tt j.j.mckenzie-smith@newcastle.ac.uk}} and D.~J.~Toms\thanks{email: {\tt d.j.toms@newcastle.ac.uk}}}

\address{Department of Physics, University of Newcastle upon Tyne,\\ Newcastle upon Tyne, NE1~7RU, U.~K.}

\date{May 2000}
\maketitle
\begin{abstract}
For the case of a relativistic scalar field at finite temperature with a chemical potential, we calculate
an exact expression for the one-loop effective action using the full fourth order determinant and
$\zeta$-function regularisation.  We find that it agrees with the exact expression for the factored
operator and thus there appears to be no mulitplicative anomaly.  The appearance of the anomaly for the fourth order
operator in the high temperature limit is explained and we show that the multiplicative anomaly can be
calculated as the difference between two $\zeta$-regularised zero-point energies.  This difference
is a result of using a charge operator in the Hamiltonian which has not been normal ordered.
\end{abstract}


\section{Introduction}
\label{sec1}

Renormalisation and regularisation techniques form a vital part of the physicist's
arsenal when performing calculations in quantum field theory.  The technique of $\zeta$-function
regularisation \cite{Ray,DowCr,Hawk} is a well established method for obtaining finite results in quantum
field theoretical calculations and has shown itself to be a very elegant and powerful method, particularly in 
curved spacetimes and spacetimes with a non-trivial topology.

Calculations involving Feynman path integrals typically involve the determinant of a differential
operator.  This determinant is an infinite product and has to be regularised in some way.  It
has been shown recently that results obtained from $\zeta$-function regularised determinants can be ambiguous 
- the source of this ambiguity is known as a multiplicative anomaly.  The multiplicative anomaly
is essentially the difference between different zeta regularised factorisations of a determinant.
For example, in calculating a determinant one may wish to factorise it in order to make calculations
easier.  Normally one would write $\det(AB)=\det(A)\det(B)$, but in the case of infinite matrices
this relation is not always correct after regularisation.  The multiplicative anomaly in a $D$
dimensional spacetime is defined as,
\begin{equation}
a_D=\ln\det(AB)-\ln\det(A)-\ln\det(B)
\label{AA}
\end{equation}
The relevance of the multiplicative anomaly for physics was first brought to light by Elizalde,
Vanzo and Zerbini \cite{Eliz,EVZ} and they showed its connection with the Wodzicki residue.
The high temperature limit of the one loop effective action for a charged scalar field with
chemical potential was considered by Elizalde, Filippi, Vanzo and Zerbini \cite{EFVZ}, and
the resulting anomaly was found to depend on the chemical potential.  It appeared therefore
that an extra term, previously overlooked, might be present in the effective action, a term
which could not be removed by renormalisation.  (This idea received criticism from Evans \cite{Evans}
and Dowker \cite{Dowk}.  Elizalde, Filippi, Vanzo and Zerbini responded to these criticisms
in \cite{R_Evans,R_Dowk}.)

It seemed that there were many different expressions for the effective action, one for each way
in which the determinant can be factorised.  So, given these varying expressions, each one
differing from another by a corresponding multiplicative anomaly, how can we know which one (if
any) is correct?  The present authors concluded in \cite{MST} that the ambiguity associated in choosing
different factorisations could only be resolved by making comparisons with calculations performed using
canonical methods.  In this paper we shall consider two factorisations; the full fourth order operator:
\begin{equation}
\Gamma_A=\frac{1}{2}\ln\det l^4\left(\left(-\Box+m^2-\mu^2\right)^2-4\mu^2\frac{\partial ^2}{\partial
t^2}\right)
\label{XA}
\end{equation}
and the `standard' factorisation of two second order operators (eg. as used in \cite{KapBook}):
\begin{equation}
\Gamma_B=-\frac{1}{2}\ln\det l^2 \left(-\Box+m^2-\mu^2+2i\mu\frac{\partial}{\partial
t}\right)-\frac{1}{2}\ln\det l^2\left(-\Box+m^2-\mu^2-2i\mu\frac{\partial}{\partial t}\right)\;.
\label{XB}
\end{equation}
In \cite{MST} an exact expression for the effective action could only be calculated for the B
factorisation, and this agreed with the standard, well known thermodynamical expression - a sum over the
zero-point energies and the thermal contributions of Bose-Einstein sums for particles and
anti-particles.
The high temperature limits of the effective action for both cases were calculated and it was found
that $\Gamma_B$ agreed with the result obtained by Haber and Weldon \cite{HW} who did not use path
integrals or $\zeta$-function regularisation. $\Gamma_A$ differed from $\Gamma_B$ by an
amount exactly equal to the multiplicative anomaly calculated in \cite{EFVZ}.  There does not seem to be 
an {\it a priori} way of determining which factorisation will yield the correct physics; an `objective'
comparison with canonical methods needs to be performed.  Clearly this is a problem if calculations using
$\zeta$-function regularisation need to be made on a system where the canonical answer is not known.

Since the high temperature expansions of $\Gamma_A$ and $\Gamma_B$ differ, it might
be thought that there would be some discrepancy between their exact expressions also.  In
Sec.~\ref{sec2} we shall calculate the exact effective action $\Gamma_A$, and show that it
actually gives the correct result, in complete agreement with $\Gamma_B$.  Is the multiplicative
anomaly therefore just an artefact of the high temperature expansion?  In Sec.~\ref{sec3} we
postulate that the multiplicative anomaly arises from the zeta-regularised zero-point energies
and is not a thermal phenomenon.  It arises in factorisations where the charge operator $Q$ has
not been normal ordered.  The multiplicative anomaly is calculated explicitly as the difference
between $\zeta$-regularised zero-point energies with and without a chemical potential.  We also
consider the interacting case.  Both calculations give multiplicative anomalies which agree
with those calculated in \cite{EFVZ}.  In Sec.~\ref{sec4} we shall draw some conclusions.

\section{The exact effective action $\Gamma_A$}
\label{sec2}

We shall use the notation and conventions of \cite{MST}, now setting $e=1$.  We are working
with a relativistic, non-interacting, charged scalar field with a chemical potential.  The
action is
\begin{eqnarray}
S[\phi] & = &\int_0^{\beta}dt\int_{\Sigma}d\sigma_x
\left(\frac{1}{2}(\dot{\phi_1} - i\mu\phi_2)^2 + 
\frac{1}{2}(\dot{\phi_2} + i\mu\phi_1)^2\right.  \nonumber \\
 & & \left. + \frac{1}{2}|\nabla\phi_1 |^2 + \frac{1}{2}|\nabla\phi_2 |^2 + \frac{1}{2}m^2(\phi_1^2+\phi_2^2)\right)\;.
\label{AB}
\end{eqnarray}
We expand about a constant background field with $\phi_1=\phi$ and $\phi_2=0$.  $\Gamma_A$
is defined formally as
\begin{eqnarray}
\Gamma_A & = & \frac{1}{2}\ln\det l^2\left(l^2S_{,ij}[\phi]\right)
\nonumber \\
 & = & \frac{1}{2}\ln\det l^4\left(\left(-\Box+m^2-\mu^2\right)^2 - 4\mu^2
\frac{\partial^2}{\partial t^2}\right)\;.
\label{AC}
\end{eqnarray}
Using $\zeta$-function regularisation \cite{Hawk} we can define:
\begin{equation}
\Gamma_A=-\frac{1}{2}\zeta'_A(0)+\frac{1}{2}\zeta_A(0)\ln l^4
\label{AD}
\end{equation}
with
\begin{equation}
\zeta_A(s)=\sum_{n}\sum_{j=-\infty}^{\infty}\left[\left(\omega_j^2-\mu^2+E_n^2\right)^2
+4\mu^2\omega_j^2\right]^{-s}
\label{AE}
\end{equation}
The $\omega_j$ are the Matsubara frequencies for scalars, $\omega_j=2\pi j/\beta$ and
$E_n^2=\sigma_n+m^2$.  $\sigma_n$ are the eigenvalues of $-\nabla^2$ on the spatial part of the manifold.  We
shall apply the Abel-Plana summation formula,
\begin{equation}
\sum_{j=-\infty}^{\infty}f(j)=\int_{-\infty}^{+\infty}f(x)dx+
\int_{-\infty+i\epsilon}^{\infty+i\epsilon}dz\left(e^{-2\pi iz}-1\right)^{-1}
\left[f(z)+f(-z)\right]
\label{AF}
\end{equation}
to evaluate $\zeta_A(s)$.  Let us label the two integrals arising on the right hand side of (\ref{AE})
$Q(s)$ and $R(s)$ respectively after using (\ref{AF}) to perform the sum over $j$.  Then,
\begin{eqnarray}
Q(s)&=&\sum_n\int_{-\infty}^{\infty}\left[\left(\left(\frac{2\pi x}{\beta}\right)^2-\mu^2+E_n^2
\right)^2+4\mu^2\left(\frac{2\pi x}{\beta}\right)^2\right]^{-s}dx \nonumber \\
&=&\sum_n\int_{-\infty}^{\infty}\left[ax^4+bx^2+c\right]^{-s}dx
\label{AG}
\end{eqnarray}
where $a=(2\pi/\beta)^4$, $b=2(2\pi/\beta)^2(E_n^2+\mu^2)$ and $c=(E_n^2-\mu^2)^2$.  Making
a simple substitution and using the identity
\begin{equation}
\alpha^{-s}=\frac{1}{\Gamma(s)}\int_0^{\infty}dt\;t^{s-1}\;e^{-\alpha t}
\label{AJ}
\end{equation}
gives us
\begin{equation}
Q(s)=\sum_n\frac{1}{\Gamma(s)}\int_0^{\infty}dt\;e^{-ct}\;t^{s-1}\int_0^{\infty}dx\;
x^{-\frac{1}{2}}\;e^{-(ax^2+bx)t}\;.
\label{AK}
\end{equation}
Evaluating the integral over $x$ first results in
\begin{equation}
Q(s)=\sum_n\frac{1}{2\Gamma(s)}\sqrt{\frac{b}{a}}\int_0^{\infty}dt\;e^{-t\left(c-\frac{b^2}{8a}\right)}
\;t^{s-1}\;K_{\frac{1}{4}}\left(\frac{b^2t}{8a}\right)
\label{AL}
\end{equation}
where $K_{\frac{1}{4}}$ is a modified Bessel function.  Performing the integral over $t$
\cite{Grad} we get,
\begin{eqnarray}
Q(s)&=&\sum_n\frac{1}{2\Gamma(s)}\sqrt{\frac{b}{a}}\frac{\sqrt{\pi}}{c^{s+1/4}}\left(
\frac{b^2}{4a}\right)^{\frac{1}{4}}\frac{\Gamma(s+\frac{1}{4})\Gamma(s-\frac{1}{4})}
{\Gamma(s+\frac{1}{2})}{}_2F_1\left(s+\frac{1}{4},\frac{3}{4};s+\frac{1}{2};1-
\frac{b^2}{4ac}\right) \nonumber \\
&=&\sum_n\sqrt{\frac{\pi}{2}}\left(\frac{\beta}{2\pi}\right)\frac{(E_n^2+\mu^2)}{(E_n^2-\mu^2)^{2s+1/2}}
\frac{\Gamma(s-\frac{1}{4})}{\Gamma(s)\Gamma(\frac{1}{4})}\int_0^1t^{s-\frac{3}{4}}(1-t)^{-\frac{3}{4}}
\left(1+\frac{4\mu^2E_n^2t}{(E_n^2-\mu^2)^2}\right)^{-\frac{3}{4}}dt
\label{AN}
\end{eqnarray}
after substituting in the values of $a$, $b$ and $c$.  This is analytic at $s=0$, and is equal to
zero, $Q(0)=0$.
It is easy to see that,
\begin{eqnarray}
Q'(0)&=&\sum_n\sqrt{\frac{\pi}{2}}\left(\frac{\beta}{2\pi}\right)\frac{(E_n^2+\mu^2)}{(E_n^2-\mu^2)^{1/2}}
\frac{\Gamma(-\frac{1}{4})}{\Gamma(\frac{1}{4})}\sqrt{2\pi}\frac{\Gamma(\frac{1}{4})}{\Gamma(\frac{3}{4})}
\left(1+\frac{4\mu^2E_n^2}{(E_n^2-\mu^2)^2}\right)^{-\frac{1}{4}}\cos\left[\frac{1}{2}\arctan
\left(\frac{4\mu^2E_n^2}{(E_n^2-\mu^2)^2}\right)^{\frac{1}{2}}\right] \nonumber \\
&=&-2\beta\sum_n(E_n^2+\mu^2)^{\frac{1}{2}}\cos\left[\frac{1}{2}\arctan\left(\frac{2\mu E_n}
{(E_n^2-\mu^2)}\right)\right] \nonumber \\
&=&-2\beta\sum_nE_n
\label{AR}
\end{eqnarray}
This last result involves the sum of zero-point energies.  Turning now to the contour integral in
(\ref{AF}), we can write
\begin{equation}
R(s)=2\sum_n\int_{-\infty+i\epsilon}^{\infty+i\epsilon}\left(e^{-2\pi iz}-1\right)^{-1}
\left[\left(\left(\frac{2\pi z}{\beta}\right)^2-\mu^2+E_n^2\right)^2+4\mu^2\left(\frac{2\pi z}{\beta}\right)^2
\right]^{-s}dz
\label{AS}
\end{equation}
since $f(z)$ is even.  There are poles at all integers on the real axis, and branch points where
the expression in square brackets in the above equation is equal to zero, namely at
$z=\pm i(\beta/2\pi)(E_n+\mu)$,$\pm i(\beta/2\pi)(E_n-\mu)$.  By taking branch cuts between the
poles in each half of the complex plane, and closing the contour in the upper half plane, it can
be shown that
\begin{equation}
R(s)=-2s\sum_n\ln\left[\left(1-e^{-\beta(E_n-\mu)}\right)\left(1-e^{-\beta(E_n+\mu)}\right)\right]
\label{AT}
\end{equation}
Hence $R(0)=0$ and
\begin{equation}
R'(0)=-2\sum_n\ln\left[\left(1-e^{-\beta(E_n-\mu)}\right)\left(1-e^{-\beta(E_n+\mu)}\right)\right]
\label{AU}
\end{equation}
Therefore $\zeta_A(0)=Q(0)+R(0)=0$ and so
\begin{eqnarray}
\Gamma_A&=&-\frac{1}{2}\zeta'_A(0) \nonumber \\
&=&\sum_n\left\{\beta E_n+\ln\left[\left(1-e^{-\beta(E_n-\mu)}\right)\left(1-e^{-\beta(E_n+\mu)}
\right)\right]\right\}
\label{AV}
\end{eqnarray}
which is the same result as was obtained for $\Gamma_B$ in \cite{MST}.  A simpler, but
perhaps less elegant way of evaluating (\ref{AE}) is presented in the appendix.

So we have a paradox - the exact expressions for $\Gamma_A$ and $\Gamma_B$
agree, while their high temperature expansions differ by the multiplicative anomaly.
The resolution of this paradox lies in the fact that the sums over the energy levels
(integrals over $k$ when $\sigma_n=k^2$) have not yet been performed in the exact
expressions we are considering.  In the high temperature expansions, the $\zeta$-functions
were expanded in powers of $\mu$, the chemical potential, and the sums over the energy
levels and Matsubara frequencies were then performed.  As we shall see in the next
section, the multiplicative anomaly arises from the chemical potential being present
in the zero-point energy contributions.  This is due to a lack of normal ordering in
the charge operator.

\section{Normal ordering and the multiplicative anomaly}
\label{sec3}

In this section we shall show that the multiplicative anomaly stems from the zero-point
energy contribution to the effective action.

\subsection{Non-interacting model}
\label{sec3.1}

There are two ways to write down the zero-point energies for the system described by
(\ref{AB}); with or without a chemical potential.  One can write $(\beta/2)\sum_n(E_n\pm\mu)$
for particles and anti-particles ($-\mu$ and $+\mu$ respectively) or one can simply write
$\beta\sum_nE_n$ (which was derived in our exact expressions for $\Gamma_A$ and $\Gamma_B$).
{\em The multiplicative anomaly is the difference between these $\zeta$-regularised zero-point
energies.}  Let us define
\begin{eqnarray}
I_{\pm}&=&\frac{\beta}{2}\sum_n(E_n\pm\mu)
\label{BA} \\
J&=&\beta\sum_nE_n\;.
\label{BB}
\end{eqnarray}
Formally of course, there is no difference between the zero-point energies in the two cases:
$I_++I_--J=0$.  But if we regularise {\em first} then the anomaly appears.  Define
\begin{eqnarray}
I_{\pm}(s)&=&\frac{\beta}{2}\sum_n(E_n\pm\mu)^{-s}
\label{BC} \\
J(s)&=&\beta\sum_nE_n^{-s}\;.
\label{BD}
\end{eqnarray}
Then, we claim that the multiplicative anomaly is
\begin{equation}
a_4=I_+(-1)+I_-(-1)-J(-1)\;.
\label{BE}
\end{equation}

Let us calculate $I_{\pm}(s)$ with $E_n^2=k^2+m^2$:
\begin{equation}
I_{\pm}(s)=\frac{\beta}{2}\frac{4\pi V}{(2\pi)^3}\int_0^{\infty}dk\;k^2\left[\left(k^2+m^2\right)
^{\frac{1}{2}}\pm\mu\right]^{-s}
\label{BF}
\end{equation}
We can binomially expand the square bracket in powers of $\mu$ up to $O(\mu^4)$.  We do not
need to consider higher order terms for reasons which will become apparent in due course.  After
performing the integrals we have,
\begin{eqnarray}
I_{\pm}(s)&=&\frac{\sqrt{\pi}\beta V}{16\pi^2}\left\{\frac{\Gamma(\frac{s}{2}-\frac{3}{2})}{\Gamma(\frac{s}{2})}
\left(m^2\right)^{\frac{3}{2}-\frac{s}{2}}\mp s\mu\frac{\Gamma(\frac{s}{2}-1)}{\Gamma(\frac{s}{2}+\frac{1}{2})}
\left(m^2\right)^{1-\frac{s}{2}}+s(s+1)\frac{\mu^2}{2}\frac{\Gamma(\frac{s}{2}-\frac{1}{2})}{\Gamma(\frac{s}{2}+1)}
\left(m^2\right)^{\frac{1}{2}-\frac{s}{2}}\right. \nonumber \\
& & \left.\mp s(s+1)(s+2)\frac{\mu^3}{6}\frac{\Gamma(\frac{s}{2})}{\Gamma(\frac{s}{2}+\frac{3}{2})}\left(m^2\right)^{-\frac{s}{2}}
+s(s+1)(s+2)(s+3)\frac{\mu^4}{24}\frac{\Gamma(\frac{s}{2}+\frac{1}{2})}{\Gamma(\frac{s}{2}+2)}\left(m^2\right)^{-\frac{s}{2}-\frac{1}{2}}
\right\}\;.
\label{BG}
\end{eqnarray}
The terms with odd powers of $\mu$ cancel when we write down an expression for $I_+(s)+I_-(s)$.
In the $\mu^2$ and $\mu^4$ terms, the $\Gamma$-function in the numerator is divergent at
$s=-1$, but can be analytically continued to cancel away the factor of $(s+1)$ multiplying
each term.  Thus,
\begin{eqnarray}
I_+(s)+I_-(s)&=&\frac{\sqrt{\pi}\beta V}{8\pi^2}\left\{\frac{\Gamma(\frac{s}{2}-\frac{3}{2})}{\Gamma(\frac{s}{2})}
\left(m^2\right)^{\frac{3}{2}-\frac{s}{2}}+2s\mu^2\frac{\Gamma(\frac{s}{2}+\frac{3}{2})}{(s-1)\Gamma(\frac{s}{2}+1)}\left(m^2\right)^{\frac{1}{2}-\frac{s}{2}}\right. \nonumber \\
& & \left.+s(s+2)(s+3)\frac{\mu^4}{12}\frac{\Gamma(\frac{s}{2}+\frac{3}{2})}{\Gamma(\frac{s}{2}+2)}
\left(m^2\right)^{-\frac{s}{2}-\frac{1}{2}}\right\}\;.
\label{BH}
\end{eqnarray}
All even, higher order terms in $\mu$ have analytic $\Gamma$-functions in the numerator, and
so the $(s+1)$ ensures they are all zero at $s=-1$.  This is why we were able to stop expanding at
fourth order in $\mu$.

Turning now to $J(s)$, we see it is exactly the first term of (\ref{BH}):
\begin{eqnarray}
J(s)&=&\frac{4\pi\beta V}{(2\pi)^3}\int_0^{\infty}dk\;k^2\left(k^2+m^2\right)^{-\frac{s}{2}} \nonumber \\
&=&\frac{\sqrt{\pi}\beta V}{8\pi^2}\frac{\Gamma(\frac{s}{2}-\frac{3}{2})}{\Gamma(\frac{s}{2})}
\left(m^2\right)^{\frac{3}{2}-\frac{s}{2}}
\label{BI}
\end{eqnarray}
Using (\ref{BE}), it is now a straightforward matter to show that the multiplicative anomaly is
\begin{equation}
a_4=\frac{\beta V}{8\pi^2}\mu^2\left(m^2-\frac{\mu^2}{3}\right)
\label{BJ}
\end{equation}
in agreement with \cite{EFVZ}.

This calculation sheds some light on why the high temperature expansions of $\Gamma_A$
and $\Gamma_B$ differ, and the exact expressions agree.  In the high temperature situation,
the integrations over $k$ were carried out after the expansions, and for a reason which is not
clear to us at the present, the chemical potentials in the zero-point energies of the
A-factorisation were not able to cancel.  So the zero-point energies were of the form
$(\beta/2)\sum_n(E_n\pm\mu)$.  In the B-factorisation, the energy levels were just $\beta\sum_nE_n$
and so there was no anomaly.  In the exact expressions for $\Gamma_A$ and $\Gamma_B$, the integrals
have not even been performed, and so the $+\mu$ and $-\mu$ simply disappear, leaving no trace of a
discrepancy.

Although simply by looking at the A factorisation (\ref{AC}), (\ref{AE}), we cannot say whether
or not it will produce a multiplicative anomaly in the high temperature expansion, given that
we know it {\em does} produce an anomaly, we can say something about {\em why} it does.  The
canonical energy levels for our system are derived from the Hamiltonian operator $H$,
\begin{equation}
H=\sum_nE_n(a_n^{\dagger}a_n+\frac{1}{2}+b_n^{\dagger}b_n+\frac{1}{2})
\label{BK}
\end{equation}
where $a_n^{\dagger}$, $a_n$ ($b_n^{\dagger}$, $b_n$) are the creation and annihilation operators
for particles (anti-particles).  For a system of charged fields with a chemical potential, the
full Hamiltonian (which is the argument of the exponential in the partition function) is
\begin{equation}
\bar{H}=H-\mu:Q:
\label{BL}
\end{equation}
where $:Q:$ is the {\em normal ordered} charge operator
\begin{equation}
:Q:=\sum_n(a_n^{\dagger}a_n-b_n^{\dagger}b_n)\;.
\label{BM}
\end{equation}
Note that we have to normal order by hand.  There is no good {\em mathematical} reason why we
normal order, we just like to have an uncharged vacuum:
\begin{equation}
\langle 0|:Q:|0\rangle=0\;.
\label{BN}
\end{equation}
It now becomes clear why we have two different expressions for the energy levels (\ref{BA}), (\ref{BB}).
They correspond to the eigenvalues of $\bar{H}$ and $H$ respectively in the case when the charge operator
$Q$ is not normal ordered.  So to avoid having an anomaly we need to ensure that both $H$ and $\bar{H}$
have the same eigenvalues - we need to normal order $Q$.  This implies that the A factorisation, which
gives rise to an anomaly, is not normal ordered. This is a symptom of using Feynman path integrals,
indeed Bernard was aware of this in 1974 - the last sentence in section II of his seminal paper
\cite{Bern} reads, `\ldots the functional-integral formalism never does normal ordering for us.'

\subsection{The interacting case}
\label{sec3.2}

The multiplicative anomaly can also be calculated in the interacting case, and can again be
seen to be the difference of $\zeta$-regularised zero-point energies.

Let us define
\begin{eqnarray}
X_{\pm}&=&\frac{\beta}{2}\sum_n\left[E_n^2+\frac{\lambda\phi^2}{3}+\mu^2\pm\left(4\mu^2\left(E_n^2
+\frac{\lambda\phi^2}{3}\right)+\frac{\lambda^2\phi^4}{36}\right)^{\frac{1}{2}}\right]^{\frac{1}{2}} \\
\label{BO}
Y&=&\frac{\beta}{2}\sum_n\left(E_n^2+\frac{\lambda\phi^2}{2}\right)^{\frac{1}{2}}+
\frac{\beta}{2}\sum_n\left(E_n^2+\frac{\lambda\phi^2}{6}\right)^{\frac{1}{2}}
\label{BP}
\end{eqnarray}
in an analogous way to $I_{\pm}$ and $J$ in the non-interacting case.  (See for example
\cite{BBD} for a full derivation of the energy levels).  Then we have
\begin{eqnarray}
X_{\pm}(s)&=&\frac{\beta}{2}\sum_n\left[E_n^2+\frac{\lambda\phi^2}{3}+\mu^2\pm\left(4\mu^2\left(E_n^2
+\frac{\lambda\phi^2}{3}\right)+\frac{\lambda^2\phi^4}{36}\right)^{\frac{1}{2}}\right]^{-\frac{s}{2}} \\
\label{BQ}
Y(s)&=&\frac{\beta}{2}\sum_n\left(E_n^2+\frac{\lambda\phi^2}{2}\right)^{-\frac{s}{2}}+
\frac{\beta}{2}\sum_n\left(E_n^2+\frac{\lambda\phi^2}{6}\right)^{-\frac{s}{2}}
\label{BR}
\end{eqnarray}
To evaluate $X_{\pm}(s)$, we expand the square root inside the square bracket in powers of
$\lambda$ up to $O(\lambda^2)$:
\begin{equation}
X_{\pm}(s)=\frac{\beta}{2}\sum_n\left[E_n^2+\frac{\lambda\phi^2}{3}+\mu^2\pm 2\mu E_n\pm
\frac{\mu\lambda\phi^2}{3E_n}\pm\frac{\lambda^2\phi^4}{144\mu E_n}\mp\frac{\mu\lambda^2\phi^4}
{36E_n^3}\right]^{-\frac{s}{2}}\;.
\label{BS}
\end{equation}
We note that we can switch from $X_+(s)$ to $X_-(s)$ by letting $\mu\rightarrow -\mu$.  Therefore
we shall work with $X_+(s)$ for simplicity, and let $\mu\rightarrow -\mu$ to write down $X_-(s)$
at the end of the calculation.  $X_+(s)$ can be written in the form,
\begin{eqnarray}
X_+(s)&=&\frac{\beta}{2}\sum_n\left[a_0+a_1\lambda+a_2\lambda^2\right]^{-\frac{s}{2}} \nonumber \\
&=&\frac{\beta}{2}\sum_n a_0^{-\frac{s}{2}}\left[1+a_0^{-1}a_1\lambda+a_0^{-1}a_2\lambda^2\right]^{-\frac{s}{2}} \nonumber \\
&=&\frac{\beta}{2}\sum_n\left(a_0^{-\frac{s}{2}}-\frac{1}{2}sa_0^{-\frac{s}{2}-1}a_1\lambda
-\frac{1}{2}sa_0^{-\frac{s}{2}-1}a_2\lambda^2+\frac{1}{8}s(s+2)a_0^{-\frac{s}{2}-2}a_1^2\lambda^2\right)
\label{BT}
\end{eqnarray}
where
\begin{eqnarray}
a_0&=&(E_n+\mu)^2 \nonumber \\
a_1&=&\frac{\phi^2}{3}\left(1+\mu E_n^{-1}\right) \nonumber \\
a_2&=&\frac{\phi^4}{36}\left(\frac{E_n^{-1}}{4\mu}-\mu E_n^{-3}\right)\;.
\label{BU}
\end{eqnarray}
It is then a straightforward (if a little tedious) matter to expand each term of (\ref{BT})
and integrate over $k$, as was done in the non-interacting case.

After the dust settles, we find
\begin{eqnarray}
X_+(s)+X_-(s)&=&\frac{\sqrt{\pi}\beta V}{8\pi^2}\left\{\frac{\Gamma(\frac{s}{2}-\frac{3}{2})}{\Gamma(\frac{s}{2})}
\left(m^2\right)^{\frac{3}{2}-\frac{s}{2}}+2s\mu^2\frac{\Gamma(\frac{s}{2}+\frac{3}{2})}{(s-1)\Gamma(\frac{s}{2}+1)}
\left(m^2\right)^{\frac{1}{2}-\frac{s}{2}}\right. \nonumber \\
& &\left.+s(s+2)(s+3)\frac{\mu^4}{12}\frac{\Gamma(\frac{s}{2}+\frac{3}{2})}{\Gamma(\frac{s}{2}+2)}
\left(m^2\right)^{-\frac{s}{2}-\frac{1}{2}}\right\}-\frac{\sqrt{\pi}\beta V\lambda\phi^2}{48\pi^2}
\left\{s\frac{\Gamma(\frac{s}{2}-\frac{1}{2})}{\Gamma(\frac{s}{2}+1)}\left(m^2\right)^{\frac{1}{2}-\frac{s}{2}}\right. \nonumber \\
& &\left.+s(s+2)\mu^2\frac{\Gamma(\frac{s}{2}+\frac{3}{2})}{\Gamma(\frac{s}{2}+2)}
\left(m^2\right)^{-\frac{1}{2}-\frac{s}{2}}\right\}+\frac{5\sqrt{\pi}\beta V\lambda^2\phi^4 
s(s+2)}{2304\pi^2}\frac{\Gamma(\frac{s}{2}+\frac{1}{2})}{\Gamma(\frac{s}{2}+2)}
\left(m^2\right)^{-\frac{1}{2}-\frac{s}{2}}
\label{BV}
\end{eqnarray}
By expanding (\ref{BR}) to order $\lambda^2$, it can easily be shown that
\begin{equation}
Y(s)=\frac{\sqrt{\pi}\beta V}{8\pi^2}\left\{\frac{\Gamma(\frac{s}{2}-\frac{3}{2})}{\Gamma(\frac{s}{2})}
\left(m^2\right)^{\frac{3}{2}-\frac{s}{2}}-s\frac{\lambda\phi^2}{6}\frac{\Gamma(\frac{s}{2}-\frac{1}{2})}{\Gamma(\frac{s}{2}+1)}
\left(m^2\right)^{\frac{1}{2}-\frac{s}{2}}+\frac{5}{288}s(s+2)\lambda^2\phi^4\frac{\Gamma(\frac{s}{2}+\frac{1}{2})}{\Gamma(\frac{s}{2}+2)}
\left(m^2\right)^{-\frac{1}{2}-\frac{s}{2}}\right\}
\label{BW}
\end{equation}
and hence,
\begin{eqnarray}
a_4&=&X_+(-1)+X_-(-1)-Y(-1) \nonumber \\
&=&\frac{\beta V}{8\pi^2}\mu^2\left(m^2+\frac{\lambda\phi^2}{3}-\frac{\mu^2}{3}\right)\;.
\label{BX}
\end{eqnarray}
This agrees with the result in \cite{EFVZ} except for a term in $\lambda^2\phi^4$.  This is of
no consequence however, since any term proportional to the background field $\phi$ (but not the
chemical potential) may be added to the effective action without changing the physics of the
system.  All such terms can be harmlessly absorbed by renormalisation.

This section has shown that the multiplicative anomaly has its roots in the manipulation of
infinite sums - the non-interacting case in particular demonstrates how the anomaly can appear
in relatively simple situations.  Elizalde showed the existence of the multiplicative anomaly
in possibly the simplest of all cases - infinite, diagonal matrices with real numbers \cite{Eliz2}.
The first worked example in \cite{Eliz2} is striking in its similarity to the calculation
presented above in the non-interacting model.

\section{Conclusions}
\label{sec4}

We have demonstrated that the multiplicative anomaly originates in the zero-point energies of
fields and is a consequence of shifting the energies by a constant amount.  When regularisation
is performed, these shifts ($+\mu$ and $-\mu$ for example) are unable to cancel and result in
a multiplicative anomaly.  It seems therefore that in order to perform anomaly-free calculations
one must resist the temptation to integrate over the momentum until the very end, after say,
a high temperature expansion has been written down.

It should be borne in mind that the functional integral approach to quantum field theory is
not as complete as the canonical one.  As was mentioned in Sec.~\ref{sec3}, path integrals
completely neglect normal ordering.  Coleman discusses the merits of path integrals at length
in his Erice lectures \cite{Cole} and echoes the comments of Bernard; the functional integral
approach does not normal order.

From inspection of equations (\ref{BA}) and (\ref{BB}) it is tempting to conclude that
$\zeta$-function regularisation may be to blame for the multiplicative anomaly (as opposed to
functional integration).  Certainly it does not seem likely that an anomaly would survive if 
say, dimensional regularisation were used to calculate the difference between (\ref{BA}) and
(\ref{BB}).  But it should be remembered that these equations were written down almost na\"{\i}vely,
to demonstrate the source of the anomaly; as was mentioned above, one should wait until the last
possible moment before performing the integration over $k$, after everything that can cancel has done
so.  Nevertheless this does not seem very satisfactory.  {\em Why} is there a difference between
(\ref{BC}) and (\ref{BD})?  The mathematical properties of the $\zeta$-function are rigorous
and well defined - to negatively criticise the whole subject of $\zeta$-function regularisation
is a step not to be taken lightly.  The paper by Elizalde \cite{Eliz2} provides some very
interesting mathematical examples of multiplicative anomalies derived from infinite matrices,
sometimes using nothing more than Riemann's $\zeta$-function.

These mathematical peculiarities aside, it has been clearly demonstrated in this paper that the problem
associated with the multiplicative anomaly can be removed by considering a Hamiltonian with a normal
ordered charge operator.  Normal ordering is in some ways an artificial procedure that physicists perform
to make the theory more physical - it is not prescribed by the theory and there is no mathematical reason
why it is done.  Using canonical techniques it is easy to see how to normal order, unfortunately it is
not so obvious in the functional integral approach - equation (\ref{XA}) is not normal ordered, but
(\ref{XB}) is.  An important step in understanding the multiplicative anomaly would be finding an {\em a
priori} method of knowing which factorisations lead to anomalies and which do not, without having to
compare with canonical calculations.

The problem of the multiplicative anomaly appears to be quite deeply rooted in the mathematics of
infinite, divergent series.

\acknowledgments

J.~J.~M-S. would like to thank Antonino Flachi for inspiring discussions.

\appendix
\section*{An alternative method of calculating $\Gamma_A^{(1)}$}

The method presented here for calculating (\ref{BE}) is simpler than that given in the main
text, but has the disadvantage that the $\zeta$-function can only be evaluated at $s=0$.  The
method in the main text can be used for an arbitrary value of $s$ (although the integral in
(\ref{AN}) may have to evaluated numerically).

We can re-write (\ref{AE}) as,
\begin{equation}
\zeta_A(s;a,b)=\left(\frac{2\pi}{\beta}\right)^{-4s}\sum_n\sum_{j=-\infty}^{\infty}
\left(j^2+a^2\right)^{-s}\left(j^2+b^2\right)^{-s}
\label{AppA}
\end{equation}
where $a=\beta/2\pi(E_n+\mu)$ and $b=\beta/2\pi(E_n-\mu)$.  Differentiating with respect to
$(a^2)$ and $(b^2)$:
\begin{eqnarray}
\frac{\partial}{\partial(a^2)}\zeta_A(s;a,b)&=&-sf(s;a,b) \nonumber \\
\frac{\partial}{\partial(b^2)}\zeta_A(s;a,b)&=&-sg(s;a,b)
\label{AppB}
\end{eqnarray}
where
\begin{eqnarray}
f(s;a,b)&=&\left(\frac{2\pi}{\beta}\right)^{-4s}\sum_n\sum_{j=-\infty}^{\infty}
\left(j^2+a^2\right)^{-s-1}\left(j^2+b^2\right)^{-s} \nonumber \\
g(s;a,b)&=&\left(\frac{2\pi}{\beta}\right)^{-4s}\sum_n\sum_{j=-\infty}^{\infty}
\left(j^2+a^2\right)^{-s}\left(j^2+b^2\right)^{-s-1}\;.
\label{AppC}
\end{eqnarray}
Clearly,
\begin{eqnarray}
\frac{\partial}{\partial(a^2)}\zeta_A(0;a,b)&=&0 \nonumber \\
\frac{\partial}{\partial(b^2)}\zeta_A(0;a,b)&=&0
\label{AppD}
\end{eqnarray}
and so we can conclude
\begin{equation}
\zeta_A(0;a,b)=\cal{C}
\label{AppE}
\end{equation}
where ${\cal C}$ is a constant independent of $a$ and $b$.  We can set $a=b$ in order
to evaluate the left hand side of (\ref{AppE}) and determine the constant ${\cal C}$.  Using the
Plana formula to calculate $\zeta_A(s;a,a)$ is straightforward:
\begin{equation}
\sum_{j=-\infty}^{\infty}\left(j^2+a^2\right)^{-2s}=\int_{-\infty}^{\infty}dj\left(j^2+a^2\right)^{-2s}
+2\int_{-\infty+i\epsilon}^{\infty+i\epsilon}\left(e^{-2\pi iz}-1\right)^{-1}\left(z^2+a^2\right)^{-2s}dz
\label{AppF}
\end{equation}
and so
\begin{equation}
\zeta_A(s;a,a)=\left(\frac{2\pi}{\beta}\right)^{-4s}\sum_n\left[\sqrt{\pi}\frac{\Gamma(2s-\frac{1}{2})}{\Gamma(2s)}
\left(a^2\right)^{\frac{1}{2}-2s}-4s\ln\left(1-e^{-2\pi a}\right)\right]
\label{AppG}
\end{equation}
(See the appendix of \cite{MST} for a more thorough evaluation of a similar sum.)  Consequently
$\zeta_A(0;a,a)={\cal C}=0$.  Next we need to evaluate $\zeta'_A(0;a,b)$,
\begin{eqnarray}
\frac{\partial}{\partial(a^2)}\zeta'_A(0;a,b)&=&-f(0;a,b) \nonumber \\
\frac{\partial}{\partial(b^2)}\zeta'_A(0;a,b)&=&-g(0;a,b)
\label{AppH}
\end{eqnarray}
The functions $f(s;a,b)$ and $g(s;a,b)$ can easily be calculated at $s=0$:
\begin{eqnarray}
\frac{\partial}{\partial(a^2)}\zeta'_A(0;a,b)&=&-\sum_n\frac{\pi}{a}\coth(\pi a) \nonumber \\
\frac{\partial}{\partial(b^2)}\zeta'_A(0;a,b)&=&-\sum_n\frac{\pi}{b}\coth(\pi b)\;.
\label{AppI}
\end{eqnarray}
Integrating,
\begin{equation}
\zeta'_A(0;a,b)=\sum_n\left\{-2\ln\left[\sinh(\pi
a)\right]-2\ln\left[\sinh(\pi b)\right]\right\}+{\cal K}\;.
\label{AppJ}
\end{equation}
Again there is a constant ${\cal K}$, independent of $a$ and $b$.  We can evaluate it in the same
way as before,
\begin{equation}
{\cal K}=\zeta'_A(0;a,a)+4\sum_n\ln\left[\sinh(\pi a)\right]\;.
\label{AppK}
\end{equation}
Using (\ref{AppG}) we see that ${\cal K}=-4\ln 2$.  Writing the $\sinh$s
in (\ref{AppJ}) as
exponentials we arrive at
\begin{equation}
\zeta'_A(0;a,b)=\sum_n\left\{-2\pi a -2\pi b -2\ln\left(1-e^{-2\pi a}\right)
-2\ln\left(1-e^{-2\pi b}\right)\right\}\;.
\label{AppL}
\end{equation}
Substituting in the values of $a$ and $b$,
\begin{equation}
\zeta'_A(0)=\sum_n\left\{-\beta(E_n+\mu)-\beta(E_n-\mu)-2\ln\left(1-e^{-\beta(E_n+\mu)}\right)
-2\ln\left(1-e^{-\beta(E_n-\mu)}\right)\right\}
\label{AppM}
\end{equation}
and so finally we have,
\begin{eqnarray}
\Gamma_A^{(1)}&=&-\frac{1}{2}\zeta'_A(0) \nonumber \\
&=&\sum_n\left\{\beta E_n+\ln\left(1-e^{-\beta(E_n+\mu)}\right)
+\ln\left(1-e^{-\beta(E_n-\mu)}\right)\right\}\;.
\label{AppN}
\end{eqnarray}
It is interesting to see that the zero-point energies in (\ref{AppM}) are written initially
with positive and negative chemical potentials.

\end{document}